\documentclass[11pt]{article}
\usepackage[dvips]{graphics}
\newcommand{\eld}{\begin{equation} \displaystyle}
\newcommand{\ee}{\end{equation}}
\newcommand{\eald}{\begin{eqnarray} \displaystyle}
\newcommand{\ea}{\end{eqnarray}}

\begin{document}

\title{Improved Gauge Actions on Anisotropic Lattices I\\
\large
    Study of Fundamental Parameters \\ in the Weak Coupling Limit }
\author{S.Sakai\\
Faculty of Education, Yamagata University, \\
Yamagata, 990-8560, Japan\\\\
T.Saito \\
Faculty of Science, Hiroshima University, \\
Higashi-Hiroshima, 739-8526, Japan \\\\
A.Nakamura\\
Research Institute of Information Sciences and Education, \\
Hiroshima University, Higashi-Hiroshima, 739-8521, Japan}

\date{\empty}
\maketitle

\begin{abstract}

On anisotropic lattices with the anisotropy $\xi=a_\sigma/a_\tau$ the
following  basic parameters are calculated by perturbative method: 
(1) the renormalization of the gauge coupling in spatial and temporal
directions, $g_\sigma$ and $g_\tau$, (2) the $\Lambda$ parameter,
(3) the ratio of the renormalized and bare anisotropy $\eta=\xi/\xi_B$
and (4) the derivatives of the coupling constants with respect to 
$\xi$, $\partial g_\sigma^{-2}/\partial \xi$ and 
$\partial g_\tau^{-2}/\partial \xi$.
We employ the improved gauge actions which consist of plaquette and 
six-link rectangular loops, $c_0 P(1 \times 1)_{\mu \nu} + c_1 P(1
\times 2)_{\mu \nu}$.
This class of actions covers Symanzik, Iwasaki and DBW2 actions.
The ratio $\eta$  shows an impressive behavior as a function
of $c_{1}$, i.e.,$\eta>1$ for the standard Wilson and Symanzik actions, 
while $\eta<1$ for Iwasaki and DBW2 actions.   This is confirmed
non-perturbatively by numerical simulations in weak coupling regions. 
The derivatives $\partial g^{-2}_{\tau}/\partial \xi$ and
$\partial g^{-2}_{\sigma}/\partial \xi$ also change sign as $-c_{1}$
increases. 
For Iwasaki and DBW2 actions they become opposite sign to those
for standard and Symanzik actions.
However, their sum is independent of the type of actions due to
Karsch's sum rule.
\end{abstract}


\section{Introduction}

Anisotropic lattices allow us to carry out numerical simulations with
the fine temporal resolution while keeping the spatial lattice spacing 
coarse, i.e. $a_\tau<a_\sigma$, where $a_\tau$ and $a_\sigma$ are
lattice spacing in the temporal and spatial directions, respectively.
This is especially important for QCD Monte Carlo simulation at finite 
temperature and heavy particle spectroscopy.
There have been many such calculations like glue
thermodynamics\cite{hashi}, hadron masses at the finite 
temperature\cite{QCDTARO1}, glueballs\cite{morning} and heavy quark
spectra\cite{klassen2}.
The anisotropic lattice may become an important tool for the
calculation of the transport
coefficients of the quark gluon plasma\cite{qgp1,qgp2} and for the
determination of spectral functions at finite
temperature\cite{spec1,spec2}.
In numerical simulations on the anisotropic lattice we need the
information upon the renormalization of anisotropy which is given by
$\eta=\xi/\xi_B$ where $\xi$ is the renormalized anisotropy,
$a_\sigma/a_\tau$, and $\xi_B$ is the bare one.
Karsch has first studied anisotropic lattices
perturbatively with the standard plaquette action and
obtained $\eta$ together with anisotropy coefficients which are defined
by the derivatives of the spatial and temporal gauge couplings with
respect to  $\xi$, i.e., $\partial g_\sigma^{-2} / \partial \xi$ and
$\partial g_\tau^{-2} / \partial \xi$, and the QCD $\Lambda$ parameter
\cite{karsch}.
In Ref.\cite{nakamura} $\eta$ was determined non-perturbatively by
analyzing Wilson loops in numerical simulations. 

Anisotropic lattices play an essential role in the analysis of
thermodynamics of QCD. To get the thermodynamics quantities
like the internal energy and the pressure
from numerical simulations, one needs to know the anisotropy coefficients.

On the anisotropic lattice, we can change the temperature by changing
not only $N_\tau$ but also $a_\tau$.
This allows us to adjust the temperature continuously with fixed
spatial volume.

Recently it has been recognized that improved actions are effective to
get reliable results from lattice QCD simulations on the relatively coarse
lattice; lattice artifacts due to the discretization are expected to be
much less. Therefore, it is important to employ improved actions in the
anisotropic lattice calculations, since the lattice is rather coarse in
spatial direction in current numerical simulations. 
Garc\'{\i}a P\'erez and van Baal pursued first this direction, i.e.,
 they have determined the one-loop correction to the anisotropy for 
the square Symanzik improved action\cite{garcia}.

In this article we study the improved actions which consist of plaquette 
and six-link rectangular loops,
\begin{equation}
 S \propto \sum \left[c_0 P(1 \times 1)_{\mu \nu} 
 + c_1  P(1 \times 2)_{\mu \nu}   \right],
\end{equation}
where $c_0$ and $c_1$ satisfy the relation $c_0 + 8c_1=1$.
This class of actions covers tree level Symanzik action without the
tadpole improvement ($c_1=-\frac{1}{12}$)\cite{symanzik,luscher},
Iwasaki action ($c_1=-0.331$) \cite{iwasaki} and DBW2 action
($c_1=-1.4088$) \cite{taro}.
They are most widely used in simulations of recent days.
For the class of actions which consists of planar loops the
anisotropic lattice can be formulated in the same way as for the
standard plaquette action.
This may not be the case for improved actions which include non-planar
loops in three or four dimensions.

In the following, we will calculate the $\Lambda$ parameter, 
$\eta= \xi/\xi_{B}$ and the anisotropy coefficients, 
in weak coupling regions mainly by perturbative calculations.

In sect. 2 we briefly review the formulation of the anisotropic
lattice with the improved actions and summarize formulae which will be
used in this paper. In sect. 3, we outline the background field method
and discuss the removal of the infrared divergence.
In sect. 4 we present results of the perturbative calculations.
The $c_1$ dependence of $\eta$ and anisotropy coefficients are studied
in detail.
Since the behavior of $\eta$ is very important for practical use, we
will study it by the numerical simulations in sect. 5.
Section 6 is devoted to concluding remarks.

\section{Anisotropic lattice with improved gauge actions}
In case of improved actions that consist of plaquette and rectangular
loops, the anisotropic lattice may be formulated in the same way as
for the standard plaquette action\cite{karsch}. The action takes the
form

\begin{equation}
  S_g = \beta_\sigma 
   \sum_x\sum_{i>j}P_{ij}+
      \beta_\tau 
   \sum_x\sum_{i\neq4}P_{4i},
\end{equation}
\noindent
where $P_{\mu\nu}$ are plaquette and six-link rectangular loop operators in $\mu$-$\nu$ plane,
\begin{equation}
  P_{\mu\nu} = c_0 P(1 \times 1)_{\mu \nu} 
 + c_1  P(1 \times 2)_{\mu \nu}, 
\end{equation}
\noindent
which are constructed of the link variable $U_{n,\mu}$ and 
\begin{equation}
\beta_\sigma = \frac{2N_c}{g^2_\sigma(a,\xi)}\xi^{-1},\hspace{0.1cm} 
\beta_\tau = \frac{2N_c}{g^2_\tau(a,\xi)}\xi .
\end{equation}
Here $g_\sigma$ and $g_\tau$ are the coupling constants in the spatial
and temporal directions, respectively.
The action can be also written with the bare anisotropy parameter
$\xi_B$ as
\begin{equation}
  S_g = \beta_{\xi}(\frac{1}{\xi_{B}}
   \sum_x\sum_{i>j}P_{ij}+
      \xi_{B}
   \sum_x\sum_{i\neq4}P_{4i})\label{bareani},
\end{equation}
where $\beta_{\xi}=2N_c/g_\xi^2= \sqrt{\beta_{\sigma} \beta_{\tau}}$ and $\xi/\xi_B=\sqrt{g_\tau^2/g_\sigma^2}$.

The weak coupling limit of the anisotropic lattice is fully discussed
in Ref.\cite{karsch}. 
Therefore, we will summarize only equations which are necessary in the
following studies. 

In the continuum limit $a_\sigma \rightarrow 0(g \rightarrow 0)$, the
lattice spacing and the coupling $g_\xi$
are related with each other by the scale parameter $\Lambda$ 
through the renormalization group relation,
\begin{equation} 
a_\sigma \Lambda(\xi) = (b_0 g_\xi^2)^{-b_1/(2b_0^2)}
                        \exp\{-1/(2b_0 g_\xi^2)\},
\end{equation} 
where $b_0$ and $b_1$ are the universal first two coefficients of
$\beta$-function,
\begin{equation}
b_0=\frac{11N_c}{48\pi^2},\quad b_1=\frac{34}{3}(\frac{N_c}{16\pi^2})^2.
\end{equation}

We calculate the effective action using the background field method 
\cite{dashen} up to one-loop order,
\begin{equation}
\begin{array}{lll}
\displaystyle 
   S_{eff} &=& \displaystyle \frac{1}{4}
                  \left( \frac{1}{g_\sigma^2(\xi)} - C_\sigma(\xi) 
                         + O(g^2) \right)
                     \sum_{i,j} F_{ij}^{2} a_\sigma^3 a_\tau\\
                  &+& \displaystyle \frac{1}{4}
                  \left( \frac{1}{g_\tau^2(\xi)} - C_\tau(\xi) 
                         + O(g^2) \right)
      \sum_{i \neq 4} ( F_{i4}^{2} + F_{4i}^{2}) a_\sigma^3 a_\tau.
\end{array}
\label{effeaction}
\end{equation}
\noindent
Effective actions with different value of the anisotropy parameter $\xi$ correspond to different regularization scheme, but they should have the same continuum limit and we require $\Delta S_{eff}=S_{eff}^{\xi\neq 1}-S_{eff}^{\xi=1}=0$. 
Then the relations are obtained,
\begin{equation}
\frac{1}{g^{2}_{\sigma}(\xi)}= 
\frac{1}{g^{2}(1)}+(C_{\sigma}(\xi)-C_{\sigma}(1))+O(g^2),
\label{9}
\end{equation}
\begin{equation}
\frac{1}{g^{2}_{\tau}(\xi)}= 
\frac{1}{g^{2}(1)}+(C_{\tau}(\xi)-C_{\tau}(1))+O(g^2).
\label{10}
\end{equation}
In the following, the deviation of the one-loop quantum correction 
from the isotropic case is often employed and written as
$\Delta C_\sigma(\xi) = C_\sigma(\xi) - C_\sigma(1)$, 
$\Delta C_\tau(\xi) = C_\tau(\xi) - C_\tau(1)$. 

Perturbatively, all fundamental parameters on the anisotropic lattice are given in terms of $C_\sigma(\xi)$ and $C_\tau(\xi)$.
The $\Lambda$ parameter on the anisotropic lattice is given by  
\begin{equation}
  \frac{\Lambda(\xi)}{\Lambda(1)}=
        \exp\left\{ -\frac{\Delta C_\sigma(\xi)
                         +\Delta C_\tau(\xi)}{4b_0}\right\}.
\label{lambda}
\end{equation}
The quantum correction for the anisotropy parameter
$\eta=\xi/\xi_B$ is written as 
\begin{equation}
   \eta(\xi,\beta) \equiv \frac{\xi}{\xi_{B}}
    = \left(\frac{g_\tau^2}{g_\sigma^2}\right)^{\frac{1}{2}}
 =1 + \frac{N_c}{\beta}\eta_1(\xi) + O(\beta^{-2}),
 \label{12}
\end{equation}
\begin{equation}
   \eta_1(\xi) = C_\sigma(\xi) - C_\tau(\xi),\label{13}
\end{equation}
\vspace{-0.5cm}
where $ \beta=2N_c/g^2 $, and $\eta_1(\xi)$ is the quantum correction
from the one-loop calculation. 
Anisotropy coefficients are given by the derivative of the $C_{\tau}(\xi)$ and $C_{\sigma}(\xi)$ with respect to $\xi$,
\begin{equation}
\frac{\partial g_\sigma^{-2}(\xi)}{\partial \xi} =  
\frac{\partial C_\sigma(\xi) }{\partial \xi},\quad
\frac{\partial g_\tau^{-2}(\xi)}{\partial \xi} = 
\frac{\partial C_\tau(\xi) }{\partial \xi}.
\end{equation}
They play an important role in QCD thermodynamics
\cite{karsch,nakamura,engels,ejiri} as will be discussed in section 4.3.

\section{Perturbative calculation of $C_{\sigma}$ and $C_{\tau}$}

\subsection{Background field method}
We calculate $C_\sigma(\xi)$ and $C_\tau(\xi)$ in one-loop order, by applying the background field method.
The background field method on the lattice is well known
\cite{hasen,dashen} and therefore here we will only outline the
method of the calculation and stress the points related to the
anisotropic lattice.
The gauge field is decomposed into a quantum field
$\alpha_\mu$ and a background one $B_\mu$ which satisfies the
classical equation of motion,
\begin{equation}
\displaystyle
 U_{n,\mu} 
  = e^{ig_{\mu} a_{\mu} \alpha_{\mu}(n) }  U^{(0)}_{n,\mu} , \hspace{0.2cm}
 U^{(0)}_{n,\mu} =  e^{i a_{\mu} B_{\mu}(n) }.
\end{equation}
A gauge fixing term is introduced as,
\begin{equation}
 S_{g.f.} = -a_\sigma^3a_\tau
\sum_{n}\mbox{Tr}(\sum_{\mu}\overline{D}_{\mu}^{(0)}\alpha_\mu(n))^2.\\
\end{equation}
Here
\begin{equation}
    \begin{array}{ll}
 D_{\mu}^{(0)}\alpha_{\kappa}
  =\frac{1}{a_{\mu}}
 (U^{(0)}_{n,\mu} \alpha_{\kappa}(n+\mu) U^{(0) \dagger}_{n,\mu}-
           \alpha_{\kappa}(n)),\\
 \overline{D}_{\mu}^{(0)}\alpha_{\kappa}
  =\frac{1}{a_{\mu}}
   (U^{(0)}_{n,-\mu} \alpha_{\kappa}(n-\mu) U^{(0) \dagger}_{n,-\mu}-
           \alpha_{\kappa}(n)).\\
\label{covad}
    \end{array}
\end{equation}

The Faddeev-Popov term resulting from the gauge fixing is 
\begin{equation}
 S_{F.P.} = -2a_\sigma^3a_\tau
\sum_{n}\sum_{\mu}\mbox{Tr}[(D_{\mu}^{(0)}\phi(n))^{\dagger}
(D_{\mu}^{(0)}\phi(n))].
\end{equation}

The total action becomes 
\begin{equation}
 S_{tot}(\alpha_\mu,\phi,B_\mu) 
    = S_g(\alpha_\mu,B_\mu) + S_{g.f.}(\alpha_\mu,B_\mu) + 
      S_{F.P.}(\phi,B_\mu),\label{totaction}
\end{equation}
\vspace{-0.5cm}
where $S_g$ is the gauge action constructed from plaquette and
six-link rectangular loops.

It is invariant under the following gauge transformation,
\begin{equation}
   \left\{\ \
     \begin{array}{ll}
      U_{n,\mu}^{(0)} &\rightarrow\ V(n)
                    U_{n,\mu}^{(0)}V^{\dagger}(n+\mu)   \\
      \alpha_\mu(n) &\rightarrow\ V(n)\alpha_\mu(n)V^{\dagger}(n)\\
       \phi(n) &\rightarrow\ V(n)\phi(n)V^{\dagger}(n),
     \end{array}
   \right.
\end{equation}
\begin{equation}
  \left\{\ \
     \begin{array}{ll}
      D_\mu \phi(n) &\rightarrow V(n)D_\mu \phi(n) V^{\dagger}(n)\\
      \overline{D}_\mu \phi(n) &\rightarrow V(n)
                        \overline{D}_\mu \phi(n) V^{\dagger}(n),\\
     \end{array}
  \right.
\end{equation}
where $V$ is an element of $SU(N_c)$.
For the calculation of the effective action in one-loop order we
expand the total action up to second order in $\alpha_{\mu}$ and
$B_{\mu}$, which we denote as $S_{tot}^{(2)}$.
With the help of Campbell-Hausdorff's formula and a relation
$V\exp(i\alpha)=\exp(iV\alpha V^{-1})V$, we split the total action
into a classical action and a bilinear term of $\alpha_\mu$ and
$\phi$,
\begin{equation}
S_{tot} = S(B_\mu) + S_{tot}^{(2)}(\alpha_\mu,\phi,B_\mu).
\end{equation}
The term linear in $B_{\mu}$ is missing because of the equation of
motion for the background field.

Thanks to the gauge invariance of the background field, it is
sufficient to calculate the coefficients of
$p_{\mu}p_{\nu}B_{\mu}B_{\nu}$ to obtain the effective action.
 For the calculation of this term, we have applied the method
explained in the appendix of
Ref.\cite{sakai}, which makes the calculation much simpler.

It is convenient to separate the action $S_{tot}^{(2)}$  into several
parts, i.e., $S_{tot}^{(2)}=S_0+S_0^{'}+\cdots+S_6^{'}+S_{F.P.}$.
$S_0, S_0^{'}, \cdots$ and $S_6^{'}$ are symbolically expressed 
as follows\cite{sakai}
\begin{equation}
   \left\{\ \
   \begin{array}{lll}
   S_0:&    &f^2 \\
   S_0^{'}:&&(\Delta \alpha)^2\\
   S_1:&    &f[B,\alpha]\\
   S_1^{'}:&&\Delta \alpha[B,\alpha]\\
   S_2:&    &[\alpha,\alpha]W\\
   S_3:&    &[B,\alpha][B,\alpha]\\
   S_4:&    &{\it f}[B,[B,\alpha]]\\
   S_5:&    &[[B,\alpha],\alpha]W\\
   S_6:&    &{\it f}^2W^2\\
   S_6^{'}:&    &\alpha W[\alpha,W],\\
   \end{array}
   \right.\label{tenkai}
\end{equation}
where $B$ and $\alpha$ represent $B_{\mu}$ and $\alpha_{\mu}$,
respectively and $W$ and $f$ are the field strength tensors of
background and quantum fields, respectively.
 $\Delta$ is a lattice derivative when we set $U^{(0)}=1$ in
Eq.(\ref{covad}).
$S_0$ and $S_0^{'}$ are the free part of the action, 
which defines gluon propagators.
$S_1$, $S_1^{'}$ and $S_2$ terms correspond to three-point diagram 
from which we construct the one-loop self-energy.
$S_3$ to $S_6^{'}$ contribute to the tadpole self-energy.
Here $S_0^{'}$ and $S_1^{'}$ result from gauge fixing terms.
The Faddeev-Popov term is the same as the previous calculations except that the anisotropy parameter $\xi$ is included \cite{hasen,dashen}.

\subsection{Effects of anisotropy parameters}

By the integration over the quantum fields, we obtain the effective
action in one-loop order.
 We carry out the integration in the momentum space. Fourier transform
of the gauge and Faddeev-Popov fields are defined as
\begin{equation}
   a_{\mu} \alpha_{\mu}(n+1/2)= \int_{-\pi}^{\pi}
   a_{\mu} \alpha_{\mu}(ka)
   \exp(i(n+1/2)ka) \prod_{\nu}\frac{d(k_{\nu}a_\nu)}{2 \pi},
\end{equation}
\begin{equation}
   a_\sigma \phi(n) = \int_{-\pi}^{\pi}
   a_\sigma \phi(ka)
   \exp( inka) \prod_{\nu}\frac{d(k_{\nu}a_\nu)}{2 \pi}.
\end{equation}
We also define Fourier transformation of the classical field in a
similar manner.

 By this Fourier transformation, the anisotropy parameters are
factorized in the action $S_{tot}^{(2)}$.
\begin{equation}
   S_{tot,\mu \nu}^{(2)}(\xi)= X_{\mu \nu} S_{tot,\mu \nu}^{(2)}(1)
   \label{xiaction}
\end{equation}
$S_{tot, \mu \nu}^{(2)}(1)$ are already given by Iwasaki and 
Sakai\cite{sakai} on the isotropic lattice and $X_{\mu \nu}$ are defined as
\begin{equation}
 X_{\mu\nu}=
\left[
\begin{array}{cccc}
   \frac{1}{\xi},\frac{1}{\xi},\frac{1}{\xi}, \xi\\
   \frac{1}{\xi},\frac{1}{\xi},\frac{1}{\xi}, \xi\\
   \frac{1}{\xi},\frac{1}{\xi},\frac{1}{\xi}, \xi\\
   \xi,\xi,\xi,\xi^3\\
\end{array}
\right].
\label{xifactor}
\end{equation}
In this way the perturbative calculation of the anisotropic lattice
becomes very systematic and transparent.

For example the free part of the anisotropic improved action $S_{0}$
is given by
\begin{equation}
S_{0} = - \frac{1}{2} \int_{k} \sum_{\mu,\nu} \alpha_{\mu}(ka)G_{\mu\nu}
\alpha_{\nu}(-ka),
\end{equation}
\begin{equation}
\left\{
\begin{array}{ccl}
G_{ii} &=& {\displaystyle \frac{1}{\xi}}\{q_{il}\hat{k}_l^2+\hat{k}_i^2\}+\xi q_{i4}\hat{k}^2_4\\
G_{44} &=& \xi q_{4l}\hat{k}^2_l + \xi^3 \hat{k}^2_4\\
G_{ij} &=& {\displaystyle \frac{1}{\xi}}\{1-q_{ij}\}\hat{k}_i\hat{k}_j\\
G_{4j} &=& \xi     \{1-q_{4j}\}\hat{k}_4\hat{k}_j\\
G_{j4} &=& G_{4j},\\
\end{array}
\right.
\end{equation}
\begin{equation}
\left\{
\begin{array}{ccl}
\hat{k}_{\mu}&=& 2 \sin\frac{1}{2}k_{\mu} a \\
q_{\mu\nu} &=& 1 - c_1(\hat{k}_{\mu}^2 + \hat{k}_{\nu}^2)\quad(\mu\neq\nu)\\
q_{\nu\nu} &=& 0,
\end{array}
\right.
\bigskip
\end{equation}
where the $\xi^3$ term in $G_{44}$ results from the gauge fixing term and
$\int_{k}$ stands for $\prod_{\mu=1}\int_{-\pi}^{\pi}dk_{\mu}a_{\mu}/2\pi$.
Note that off-diagonal elements of $G_{\mu\nu}$ vanish for the Wilson action.

Propagators $D_{\mu\nu}$ are defined by
\begin{equation}
    <\alpha_\mu^i(k a)\alpha_\nu^j(k^{'}a)> = \delta_{ij}(2\pi)^4
                                           \delta^{(4)}(k a+k^{'}a)
                                           D_{\mu\nu}(ka ),
\end{equation}
and they are obtained by solving the equations,
\begin{equation}
G_{\mu\rho}D_{\rho\nu} = \delta_{\mu\nu}.
\end{equation}

The Faddeev-Popov propagator is given by
\begin{equation}
      D_{F.P.}(ka) = \frac{\xi}{\hat{k_i}^2 + \xi^2\hat{k_4}^2}.
\end{equation}

The explicit forms of the $S_{1}$ to $S_{6}^{'}$ terms in
Eq.(\ref{tenkai}) are obtained 
by setting $c_2=c_3=0$ in formulae of Ref.\cite{sakai} and by
introducing the anisotropy factors as shown 
in Eqs. (\ref{xiaction}) and (\ref{xifactor}).

The effective action is obtained by integrating 
$S_{tot}^{(2)}(\alpha_\mu,\phi,B_\mu)$ over $a_{\mu} \alpha_{\mu}$ and Faddeev-Popov fields $a_\sigma\phi$,
\begin{equation}
\displaystyle
   e^{-S(a_{\mu}B_\mu)} \int \mathcal{D} (a_{\mu}\alpha_{\mu})
                             \mathcal{D} (a_{\sigma}\phi)
          e^{-S_{tot}^{(2)}(a_{\mu}\alpha_{\mu},a_{\sigma} \phi,a_{\mu}B_\mu)}
    = e^{-S_{eff}(a_{\mu}B_\mu)}.
\end{equation}
Then $C_\sigma(\xi)$ and $C_\tau(\xi)$ in Eq.(\ref{effeaction}) are
obtained as coefficients of $F_{ij}^2$ and $F_{i4}^2$ in $S_{eff.}$,
respectively.
\subsection{Infrared divergence}
We will discuss here a subtle point concerning the cancellation of the infrared divergence.
The contributions for $C_{\sigma}$ and $C_{\tau}$ from the self-energy type diagram of the term $ <(S_{1}+S_{2})^2>$ have the infrared divergence. 
However in the difference given by Eqs. (\ref{9}) and (\ref{10}) they are canceled. 
But numerically the calculation of the divergent integral is very delicate problem. 
In the numerical evaluation, we discretize the momentum integration $\int d^4k$,
and the divergence comes from the segment including $k=0$.
We should not take the difference of integration with different $\xi$ directly,
because their measures of segment are different for isotropic and anisotropic lattices.
We use the following method for the calculation of the Eqs. (\ref{9}) and (\ref{10}),
\begin{equation}
\begin{array}{lll}
\displaystyle 
   C^{Imp.}_{\sigma}(\xi)-C^{Imp.}_{\sigma}(1)
           &=&(C^{Imp.}_{\sigma}(\xi)- C^{Stand.}_{\sigma}(\xi))\\
           &+&(C^{Stand.}_{\sigma}(\xi)-C^{Stand.}_{\sigma}(1))\\
           &+&(C^{Stand.}_{\sigma}(1)-C^{Imp.}_{\sigma}(1)).\\
\label{26}
\end{array}
\end{equation}
where $C^{Imp.}$ and $C^{Stand.}$  are the coefficients with improved and standard actions, respectively.

In the first term of r.h.s of Eq.(\ref{26}), $C^{Imp.}_{\sigma}(\xi)$ and $C^{Stand.}_{\sigma}(\xi)$ have the same infrared divergence and they are canceled exactly by each other.
The second term of Eq.(\ref{26}) can be calculated by the analytic integration of the $4th$ component of the loop momentum\cite{karsch}.
 The results do not include the infrared divergence and the numerical integration is stable. 
The divergence in the last term has been already calculated in Ref.\cite{sakai}.
 We have checked that our calculations for the second and the third
terms of Eq.(\ref{26}) coincide with those for Wilson and Symanzik
case respectively given in Ref.\cite{karsch} and Ref.\cite{sakai}
\footnote{Although the values of $C^{Stand.}(\xi)-C^{Stand.}(1)$ were not given in Ref.\cite{karsch}, we have calculated them from Table and formulae in Ref.\cite{karsch}}.

In this way the difference $C^{Imp.}_{\sigma}(\xi)-C^{Imp.}_{\sigma}(1)$ is calculated in numerically stable manner. 
Similar calculation has been done for $C^{Imp.}_{\tau}(\xi)-C^{Imp.}_{\tau}(1)$.

\section{Results of one-loop calculation}

Values of $C_\sigma(\xi)-C_\sigma(1)$ and $C_\tau(\xi)-C_\tau(1)$ are given in Table \ref{csct} for Symanzik, Iwasaki and DBW2.
$\xi$ is varied from 1 to 6, since these anisotropy
parameters are often used in Monte Carlo simulations on anisotropic
lattices.
%
%
\renewcommand{\arraystretch}{0.9}
\begin{table}[h]
\begin{tabular}{|c|c|c|c|c|c|c|c|}
     \hline
     & & \multicolumn{2}{|c|}{Symanzik action} &
     \multicolumn{2}{|c|}{Iwasaki action} &
     \multicolumn{2}{|c|}{DBW2 action} \\ 
     \hline
     &$\xi$&$\Delta C_\sigma$&$\Delta C_\tau$& 
     $\Delta C_\sigma$&$\Delta C_\tau$ &$\Delta C_\sigma$
     &$\Delta C_\tau$ \\
     \hline
     & 1.0 &0.00000& 0.00000& 0.00000&0.00000& 0.00000&0.00000\\
     & 1.1 &0.00541&-0.00092&-0.00201&0.00614&-0.01124&0.01464\\
     & 1.5 &0.02072&-0.00173&-0.00908&0.02327&-0.04875&0.05150\\
SU(2)& 2.0 &0.03180&-0.00085&-0.01564&0.03612&-0.08400&0.07510\\
     & 3.0 &0.04304& 0.00103&-0.02406&0.05016&-0.13384&0.09537\\
     & 4.0 &0.04862& 0.00159&-0.02934&0.05613&-0.16932&0.09766\\
     & 5.0 &0.05168& 0.00059&-0.03340&0.05692&-0.19695&0.09055\\
     & 6.0 &0.05314&-0.00162&-0.03725&0.05444&-0.21970&0.08000\\
     \hline
     & 1.0 &0.00000& 0.00000& 0.00000&0.00000& 0.00000&0.00000\\
     & 1.1 &0.00930&-0.00253&-0.00427&0.01045&-0.02087&0.02584\\
     & 1.5 &0.03602&-0.00701&-0.01852&0.03967&-0.08983&0.09206\\
SU(3)& 2.0 &0.05571&-0.00829&-0.03119&0.06184&-0.15389&0.13609\\
     & 3.0 &0.07596&-0.00856&-0.04693&0.08679&-0.24348&0.17729\\
     & 4.0 &0.08614&-0.00951&-0.05644&0.09843&-0.30658&0.18776\\
     & 5.0 &0.09182&-0.01216&-0.06350&0.10163&-0.35533&0.18221\\
     & 6.0 &0.09462&-0.01625&-0.07012&0.09953&-0.39532&0.17034\\
     \hline
\end{tabular}
\vspace{0.5cm}
\caption{$C_\sigma$ and $C_\tau$ for Symanzik, Iwasaki and DBW2 actions.
         Here $\Delta C_\sigma=C_\sigma(\xi)-C_\sigma(1)$ and
         $\Delta C_\tau=C_\tau(\xi)-C_\tau(1)$. \label{csct} }
\end{table}
\subsection{The $\Lambda$ ratio}
When we calculate physical quantities, we must take into account the 
variation of the scale $a_{\sigma}$ due to $\Lambda(\xi)$. In weak 
coupling regions it is given by Eq.(\ref{lambda}).
The $\Lambda$ ratio is calculated as a product of three factors,
\begin{equation}
    \frac{\Lambda_{Imp.}(\xi)}{\Lambda_{Imp.}(1)} =
    \frac{\Lambda_{Imp.}(\xi)}{\Lambda_{Stand.}(\xi)} \times
    \frac{\Lambda_{Stand.}(\xi)}{\Lambda_{Stand.}(1)} \times
    \frac{\Lambda_{Stand.}(1)}{\Lambda_{Imp.}(1)}.
\end{equation}
The numerical results are given in the Table \ref{lambda2} for Symanzik, Iwasaki and DBW2 and are shown in Fig. 1. 
\begin{center}
\renewcommand{\arraystretch}{0.9}
\begin{table}[h]
\begin{tabular}{|c|c|c|c|c|c|c|c|c|c|c|}
     \hline
     & & \multicolumn{3}{|c|}{$\Lambda_{Imp.}(\xi)/\Lambda_{Imp.}(1)$}
     & & \multicolumn{3}{|c|}{$\Lambda_{Imp.}(\xi)/\Lambda_{Imp.}(1)$}\\
     \hline
     &$\xi$&Symanzik&Iwasaki&DBW2&     &Symanzik&Iwasaki&DBW2\\
     \hline
     &1.0&1.00000&1.00000&1.00000&     &1.00000&1.00000&1.00000\\
     &1.1&0.97614&0.97803&0.98187&     &0.97599&0.97809&0.98234\\
     &1.5&0.90282&0.92641&0.98529&     &0.90111&0.92691&0.99204\\
SU(2)&2.0&0.84653&0.89562&1.04907&SU(3)&0.84350&0.89583&1.06597\\
     &3.0&0.78873&0.86890&1.23013&     &0.78512&0.86670&1.26812\\
     &4.0&0.76309&0.86566&1.47071&     &0.75955&0.86010&1.53177\\
     &5.0&0.75471&0.88107&1.77320&     &0.75133&0.87209&1.86138\\
     &6.0&0.75778&0.91159&2.12140&     &0.75483&0.89981&2.24211\\
     \hline
\end{tabular}
\vspace{0.5cm}
\caption{$\Lambda$ parameter ratio for improved actions.
         \label{lambda2}}
\end{table}
\end{center}
\begin{figure}[h]
\begin{center}
 \scalebox{0.4}{ \rotatebox{-90} { \includegraphics{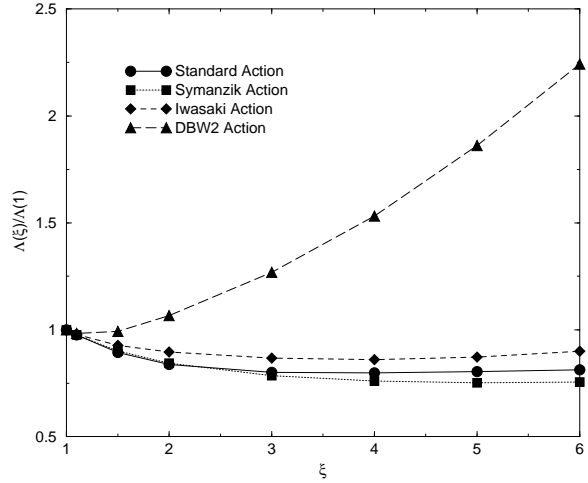} } }
\caption{The ratio of $\Lambda$ parameter for SU(3). \label{lambdafig} }
\end{center}
\end{figure}
\subsection{The $\eta$ parameter}
\begin{figure}[h] \begin{center} \scalebox{0.6}{ \rotatebox{-90}{ \includegraphics{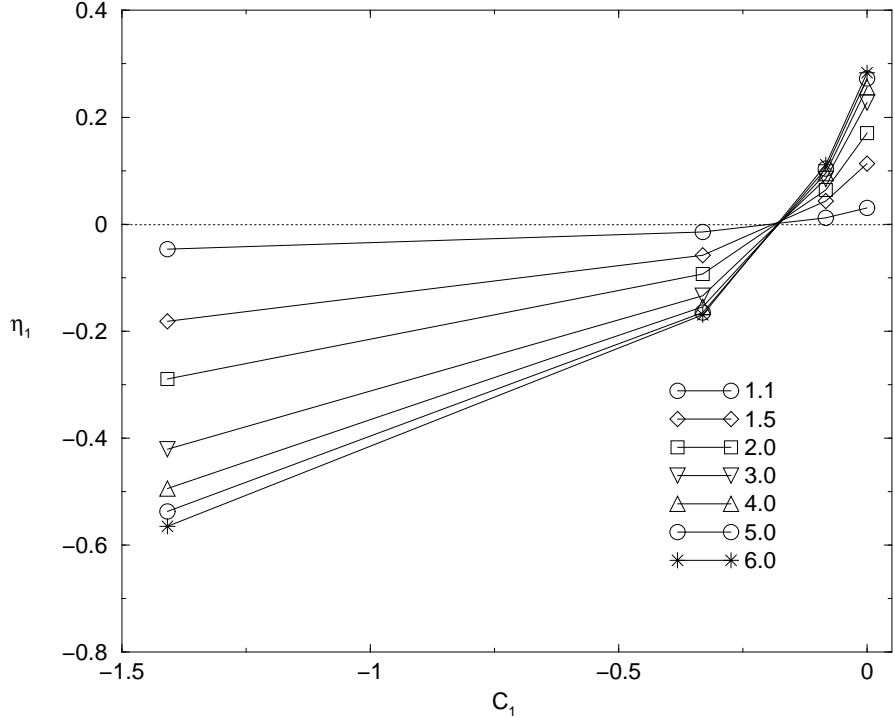} } }
\caption{The $\eta_1$ as a function of $c_1$ for SU(3).
\label{eta1c1} }
\end{center}
\end{figure}

The $\eta$ parameter is given by Eqs. (\ref{12}) and (\ref{13}). 
Because of the gauge invariance 
we can see $C_\sigma(1)=C_\tau(1)$ at the isotropic case,  
and therefore these coefficients do not appear in the definition of $\eta$.
But to cancel the infrared divergence we calculate Eq.(13) as 
$\eta_1(\xi)=(C_\sigma(\xi)-C_\sigma(1))-(C_\tau(\xi)-C_\tau(1) )$.
We plot $\eta_1$ as a function of $c_{1}$ in Fig. \ref{eta1c1}.
\begin{center}
\renewcommand{\arraystretch}{0.9}
\begin{table}[h]
\begin{tabular}{|c|c|c|c|c|c|c|c|c|c|}
     \hline
     &$\xi$&Symanzik&Iwasaki&DBW2& &Symanzik&Iwasaki&DBW2\\
     \hline
     &1.0&0.00000& 0.00000& 0.00000&     &0.00000& 0.00000& 0.00000\\
     &1.1&0.00633&-0.00816&-0.02588&     &0.01183&-0.01472&-0.04671\\
     &1.5&0.02246&-0.03235&-0.10025&     &0.04303&-0.05820&-0.18190\\
SU(2)&2.0&0.03266&-0.05177&-0.15911&SU(3)&0.06400&-0.09304&-0.28998\\
     &3.0&0.04201&-0.07423&-0.22921&     &0.08452&-0.13372&-0.42077\\
     &4.0&0.04702&-0.08548&-0.26698&     &0.09565&-0.15487&-0.49434\\
     &5.0&0.05109&-0.09032&-0.28751&     &0.10398&-0.16513&-0.53755\\
     &6.0&0.05476&-0.09170&-0.29971&     &0.11088&-0.16966&-0.56567\\
     \hline
\end{tabular}
\vspace{0.5cm}
\caption{The $\eta_1$ for improved actions.  \label{eta}}
\end{table}
\end{center}

The parameter $\eta_1$ changes sign around $c_{1}=-0.18\sim-0.19$. 
This means that in weak coupling regions, there is no renormalization for the anisotropy parameter $\xi$ for this action. 
The interesting point is that the $\beta$ dependence of the $\eta$ with Iwasaki and DBW2 actions is opposite to those with standard and Symanzik actions; As $\beta$ decreases $\eta$ decreases for Iwasaki and DBW2 action, while it increases for standard and Symanzik actions.
This is a new feature and we shall confirm it non-perturbatively in
the next section.
 
\subsection{Anisotropy coefficients}
The anisotropy coefficients, which are the derivatives of spatial and
temporal gauge couplings with respect to the anisotropy $\xi$, are
calculated as,
\begin{equation}
\frac{\partial g_\sigma^{-2}}{\partial \xi} = 
\frac{\partial}{\partial \xi} ( C_\sigma^{Imp.}(\xi)-C_\sigma^{Stand.}(\xi ) )
+\frac{\partial}{\partial \xi} ( C_\sigma^{Stand.}(\xi)-C_\sigma^{Stand.}(1) ).
\label{sumruleo}
\end{equation}
In this manner, we are free from the infrared divergence, and the
numerical evaluation is stable as in Sect. 3.3.

\begin{figure}[h]
\begin{center}
\scalebox{0.5}{ \rotatebox{-90}{\includegraphics{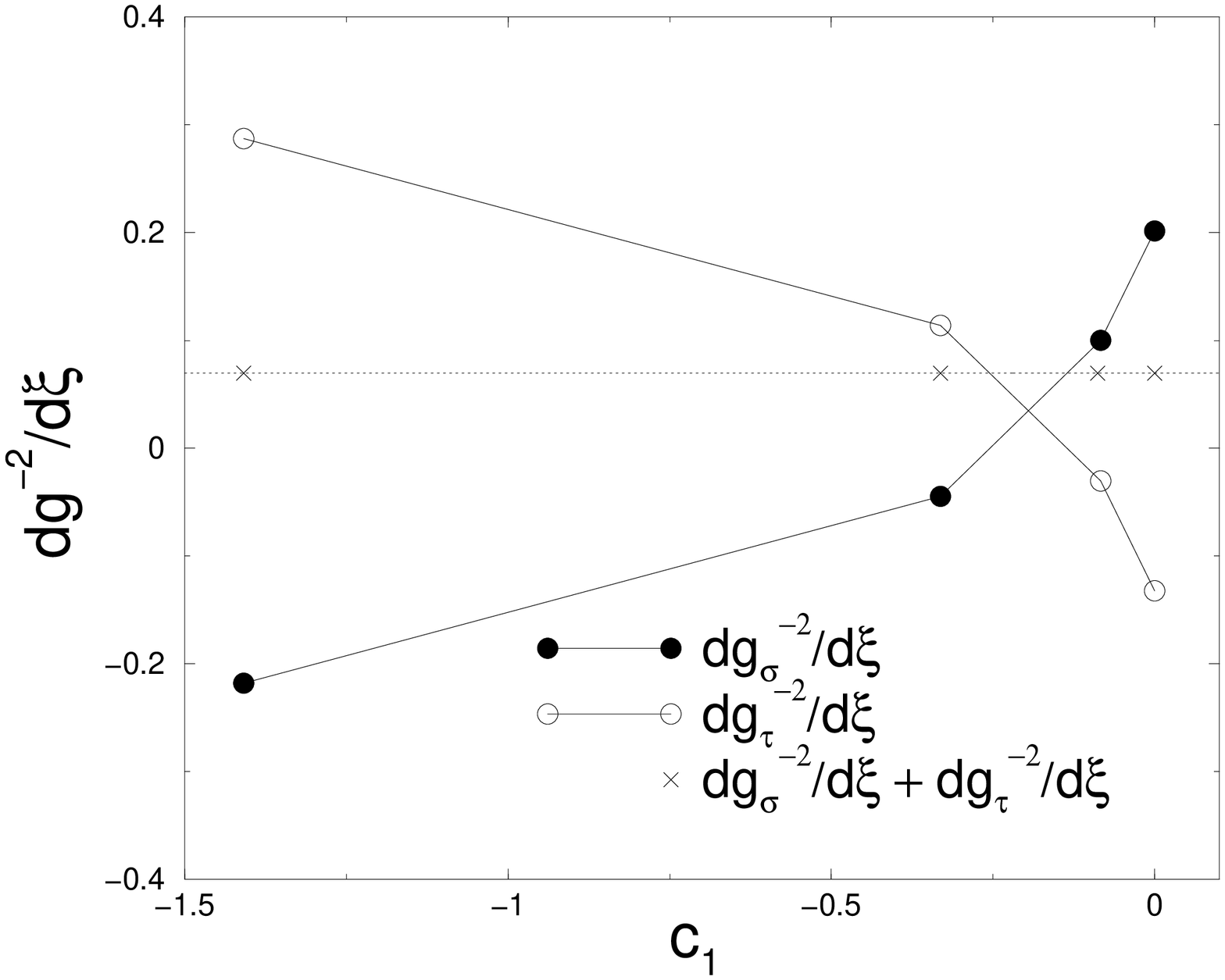}}}
\caption{ Anisotropy coefficients with the standard and improved actions for SU(3). Dotted line represents the r.h.s of Eq.(\ref{sumrule}), which
is about 0.0697 for SU(3) case.
 The cross symbol stands for the l.h.s of Eq.(\ref{sumrule}).
\label{anicoefig} }
\end{center}
\end{figure}

From the invariance of the string tension on the isotropic and anisotropic lattice, Karsch has derived the following sum rule\cite{karsch},
\begin{equation}
    \frac{\partial g_\sigma^{-2}}{\partial \xi} +
    \frac{\partial g_\tau^{-2}}{\partial \xi}   =
    \frac{11N_c}{48 \pi^2}\ \ (\xi=1,a\rightarrow0) \label{sumrule}.
\end{equation}
The same arguments are applied to improved actions.
We show results in Table \ref{anicoe} and Fig. \ref{anicoefig}.   
This sum rule is satisfied quite well.

\renewcommand{\arraystretch}{1.0}
\begin{table}[b]
\begin{tabular}{|c|c|c|c|c|}
      \hline
      &\multicolumn{2}{c|}{SU(2)}&\multicolumn{2}{c|}{SU(3)}\\
      \hline
      &{$\partial g_\sigma^{-2}/\partial \xi |_{\xi=1}$}
      &{$\partial g_\tau^{-2}\partial \xi |_{\xi=1}$}
      &{$\partial g_\sigma^{-2}/\partial \xi |_{\xi=1}$}
      &{$\partial g_\tau^{-2}/\partial \xi |_{\xi=1}$}\\
      \hline
Symanzik& 0.058406&-0.01196& 0.10025&-0.030365\\
Iwasaki &-0.020569& 0.067009&-0.044305& 0.113965\\
DBW2    &-0.117103& 0.163543&-0.21793& 0.287592\\
      \hline
\end{tabular}
\vspace{0.5cm}
\caption{The anisotropy coefficients for improved actions. 
The results are checked by estimating the
derivative numerically by using the results with $\xi=1.05, 1.1$.
\label{anicoe} }
\end{table}

An interesting point is that individual terms $\partial g_\sigma^{-2}/\partial 
\xi$ and $\partial g_\tau^{-2}/\partial \xi$ also change sign as 
$-c_{1}$ increases.  
For Iwasaki and DBW2 actions, $\partial g_\tau^{-2}/\partial \xi$
becomes positive while it is negative for standard and Symanzik
actions.
These anisotropy coefficients have a contribution to QCD thermodynamics.
For example,
Okamoto et al.\cite{okamoto} used our perturbation results to study the
energy and pressure with Iwasaki action and had no negative pressure problem,
 contrary to Wilson action case\cite{svetitsky}.

\section{Numerical results in weak coupling regions}
In the previous section we have found by the perturbative calculation
that the ratio of the renormalized and bare anisotropy, $\eta$,
becomes less than one for Iwasaki and DBW2 actions.

Since the $\eta$ is important in QCD simulations on the anisotropic
lattice, we will study its behavior further by numerical simulations.
 
Numerically the $\eta$ parameter is calculated from the relation \cite{nakamura,klassen}, 
\begin{equation}
 \eta = \xi/\xi_{B}.
\end{equation}
The anisotropy $\xi_{B}$ appears in the action given by Eq.(\ref{bareani}) while the renormalized anisotropy $\xi$ is defined by
\begin{equation}
  \xi= a_{\sigma}/a_{\tau}.
\end{equation}
For the probe of the scale in the space and temperature direction, 
we use the lattice potential in these directions, which is defined by
\begin{equation}
  V_{\sigma\tau}(\xi_{B},l,t)=\ln(\frac{W_{\sigma\tau}(l,t)}{W_{\sigma\tau}(l+1,t)}),
\end{equation}
where $W_{\sigma\tau}(l,t)$ is the Wilson loop of the size $l \times t$ in the temporal plane.
Similar formula holds for the potential in space direction.
We fix $\xi=2$, and calculate the ratio at a few $\xi_{B}$ points,
\begin{equation}
  R(\xi_{B},l,r)=\frac{V_{\sigma\sigma}(\xi_{B},l,r)}{V_{\sigma\tau}(\xi_{B},l,\xi t)}.
\end{equation}
Then we search for the point $R=1$ by interpolating $\xi_{B}$\cite{nakamura,klassen}.
In Ref.\cite{engels2}, an extensive study is done for the determination
of $\xi_B$.

\begin{figure}[h]
\begin{center}
 \scalebox{0.5}{ \rotatebox{-90} { \includegraphics{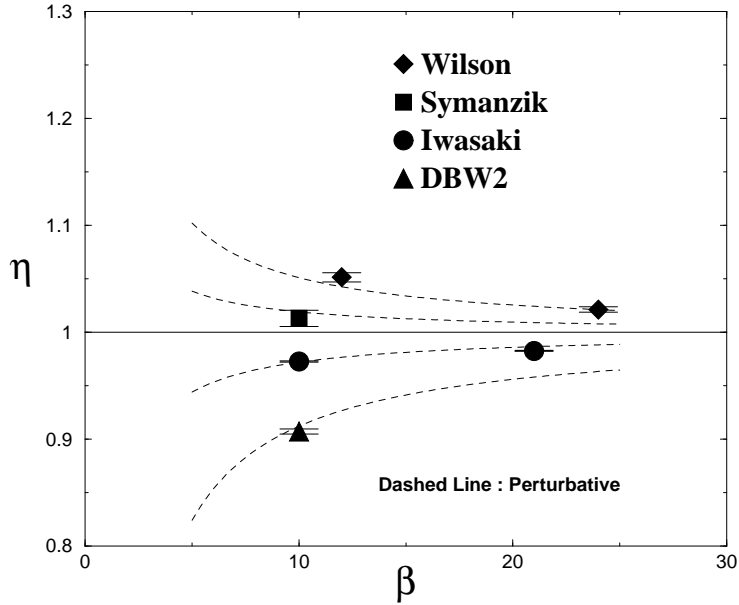} }
               }
\caption{ Perturbative and non-perturbative results of $\eta$ as a function 
of $\beta$.  
Data for Wilson actions are taken from Ref.\cite{klassen} \label{b10} }
\end{center}
\end{figure}

The simulations are done on the $12^{3} \times 24 $ lattice.
Numerical results with large $\beta$ region( $\beta \geq 10$) together
with perturbative ones are shown in Fig. \ref{b10}. 
They agree with each other at large $\beta$ region.
The $\eta$ parameter decreases as $\beta$ decreases for Iwasaki and DBW2 while it increases for standard and Symanzik actions.
 
\section{Concluding remarks}

We have calculated the QCD scale parameter $\Lambda$, which is shown in Table \ref{lambda2} and Fig. \ref{lambdafig}.
Ratios $\Lambda(\xi)/\Lambda(1)$ for standard, Symanzik and Iwasaki actions have  very similar behavior; they are slightly less than one, but the behavior of $\Lambda$ ratio of the DBW2 is quite different from those of other actions. 
DBW2 action is expected to be very near to the renormalized trajectory\cite{taro}, and may have a special feature.
 It may be interesting to study the $c_{1}$ dependence of the $\Lambda$ parameters between Iwasaki and DBW2 actions in more detail. 

The $\eta$ parameters and anisotropy coefficients, the derivatives of the coupling constants with respect to the anisotropy parameter $\xi$ are calculated in one-loop order for improved actions.
We have found that the $\eta_1$ in the Eq.(\ref{13}) changes sign for $c_1 \simeq -0.18 \sim -0.19$ as shown in Fig. \ref{eta1c1}.
The $\eta_1(\xi)$ is positive for the standard plaquette and Symanzik actions while it becomes negative for Iwasaki and DBW2 actions.
This behavior is confirmed non-perturbatively by the numerical simulations in weak coupling regions as shown in Fig. \ref{b10}.
 
We have also found that the anisotropy coefficients change sign
as $-c_1$ increases.
For Iwasaki and DBW2 actions, $\partial g^{-2}_{\tau}/\partial \xi$ is
positive while for standard and Symanzik actions it is negative.
These may be good properties for the thermodynamics using those
improved actions\cite{okamoto}.

We have found that $\eta$ obtained by the perturbation calculations is
close to one in the region $c_1\simeq -0.18 \sim -0.19$ for $\xi = 1
\sim 6 $
\footnote{The value is sligtly shifted from that in Ref.\cite{aia}.
After lattice'99, we have done an extensive study of the numerical
calculation of the loop integral, so that we can safely extrapolate
to the continuum integral.}.
A natural question is whether this is true also at intermediate and
strong coupling regions.
 Parts of the results on the $\eta$ parameter are reported at lattice
'99 at Pisa\cite{aia}.
 The study of the lattice spacing $a$ on the anisotropic lattice in
intermediate coupling regions has been started.
 Detailed results will be reported in the forthcoming paper.
 Moreover, with these fundamental properties of the improved actions on anisotropic lattices, we are going to start simulations of heavy quark spectroscopy, transport coefficients of quark gluon plasma etc. on these lattices.

\begin{center}
\bf Acknowledgments
\end{center}
 We are grateful to R. V. Gavai for the discussion for thermodynamics quantities.
This work is supported by the Grant-in-Aide for Scientific Research by Monbusho, Japan (No. 10640272). 
In this work, QCD Monte Carlo simulations have been done on SX-4 at RCNP, Osaka Univ. and
on VPP500 at KEK(High Energy Accelerator Research Organization)
 and Tsukuba Univ..
\begin{center}
\bf Appendix
\end{center}
In this appendix we show all the data used in Eq.(\ref{26}).
In Table \ref{csct1} the data of the first term in Eq.(\ref{26})
are summarized, and in Table \ref{csct2} we report the data for the standard action.
The values of the third terms in Eq.(\ref{26}) are given by isotropy cases in Table \ref{csct1}.
\renewcommand{\arraystretch}{0.9}
\begin{table}[h]
\begin{tabular}{|c|c|c|c|c|c|c|c|}
     \hline
     & & \multicolumn{2}{|c|}{Symanzik action} &
     \multicolumn{2}{|c|}{Iwasaki action} &
     \multicolumn{2}{|c|}{DBW2 action} \\
     \hline
     &$\xi$&$\delta C_\sigma(\xi)$&$\delta C_\tau(\xi)$&
     $\delta C_\sigma(\xi)$&$\delta C_\tau(\xi)$ &
     $\delta C_\sigma(\xi)$&$\delta C_\tau(\xi)$ \\
     \hline
     & 1.0 &-0.13173&-0.13173&-0.32668&-0.32668&-0.57505&-0.57505\\
     & 1.1 &-0.13690&-0.12672&-0.33928&-0.31461&-0.59687&-0.55447\\
     & 1.5 &-0.15149&-0.11355&-0.37625&-0.28349&-0.66429&-0.50362\\
SU(2)& 2.0 &-0.16130&-0.10393&-0.40371&-0.26190&-0.72042&-0.47128\\
     & 3.0 &-0.16908&-0.09261&-0.43115&-0.23844&-0.78929&-0.44160\\
     & 4.0 &-0.17135&-0.08643&-0.44427&-0.22685&-0.83262&-0.43368\\
     & 5.0 &-0.17218&-0.08359&-0.45223&-0.22221&-0.86414&-0.43694\\
     & 6.0 &-0.17292&-0.08300&-0.45828&-0.22188&-0.88909&-0.44468\\
     \hline
     & 1.0 &-0.23211&-0.23211&-0.57349&-0.57349&-0.99987&-0.99987\\
     & 1.1 &-0.24156&-0.22293&-0.59652&-0.55132&-1.03950&-0.96232\\
     & 1.5 &-0.26836&-0.19832&-0.66429&-0.49302&-1.16198&-0.86701\\
SU(3)& 2.0 &-0.28649&-0.17988&-0.71477&-0.45112&-1.26386&-0.80326\\
     & 3.0 &-0.30099&-0.15771&-0.76527&-0.40373&-1.38820&-0.73962\\
     & 4.0 &-0.30524&-0.14509&-0.78921&-0.37852&-1.46573&-0.71558\\
     & 5.0 &-0.30674&-0.13848&-0.80344&-0.36606&-1.52166&-0.71187\\
     & 6.0 &-0.30799&-0.13586&-0.81412&-0.36144&-1.56570&-0.71702\\
     \hline
\end{tabular}
\vspace{0.5cm}
\caption{$C_\sigma$ and $C_\tau$ for Symanzik, Iwasaki and DBW2 actions.
         Here 
$\delta C_\sigma(\xi)=C_\sigma^{Imp.}(\xi)-C_\sigma^{Stand.}(\xi$) and
$\delta C_\tau(\xi)=C_\tau^{Imp.}(\xi)-C_\tau^{Stand.}(\xi$).
         \label{csct1} }
\end{table}

\begin{center}
\renewcommand{\arraystretch}{0.9}
\begin{table}[h]
\begin{tabular}{|c|c|c|c|c|c|c|c|c|c|c|}
     \hline
     &$\xi$&$\Delta C_\sigma(\xi)$&$\Delta C_\tau(\xi)$& 
     &$\Delta C_\sigma(\xi)$&$\Delta C_\tau(\xi)$\\
     \hline
     &1.0&0.00000& 0.00000&     &0.00000& 0.00000\\
     &1.1&0.01058&-0.00593&     &0.01875&-0.01171\\
     &1.5&0.04048&-0.01991&     &0.07227&-0.04079\\
SU(2)&2.0&0.06137&-0.02865&SU(3)&0.11009&-0.06051\\
     &3.0&0.08039&-0.03807&     &0.14485&-0.08296\\
     &4.0&0.08824&-0.04370&     &0.15928&-0.09652\\
     &5.0&0.09213&-0.04754&     &0.16645&-0.10578\\
     &6.0&0.09433&-0.05035&     &0.17051&-0.11250\\
     \hline
\end{tabular}
\vspace{0.5cm}
\caption{$C_\sigma$ and $C_\tau$ for the standard action.
$\Delta C_\sigma(\xi)=C_\sigma^{Stand.}(\xi)-C_\sigma^{Stand.}(1)$ and
$\Delta C_\tau(\xi)=C_\tau^{Stand.}(\xi)-C_\tau^{Stand.}(1)$.
         We have evaluated these data from Table and formulae 
         in Ref.\cite{karsch}.
         \label{csct2}}
\end{table}
\end{center}

\newpage

\end{document}